\documentclass[prl,twocolumn,aps,amsmath,amssymb,nofootinbib,preprintnumbers]{revtex4-1}
\usepackage{graphicx,epsfig}
\usepackage{bm}
\usepackage{amsmath}
\usepackage{amsfonts}
\usepackage{verbatim}
 
\newcommand{\beq}{\begin{equation}}
\newcommand{\eeq}{\end{equation}}
\newcommand{\bea}{\begin{eqnarray}}
\newcommand{\eea}{\end{eqnarray}}
\def \phicmb{\phi_{\rm CMB}}

\begin{document}

\title{Observable tensor-to-scalar ratio and secondary gravitational wave background }

\author{Arindam Chatterjee$^{1}$}
\author{Anupam Mazumdar$^{2,3}$}

\affiliation{$^{1}$~ Indian Statistical Institute, 203, B.T. Road, Kolkata- 700108, India} 
\affiliation{$^{2}$~Van Swinderen Institute, University of Groningen, 9747 AG, Groningen, The Netherlands}
\affiliation{$^{3}$~Kapteyn Astronomical Institute, University of Groningen, 9700 AV Groningen, The Netherlands}


\begin{abstract}
In this paper we will highlight how a simple vacuum energy dominated {\it inflection-point} inflation can match the 
current data from cosmic microwave background radiation, and predict large primordial tensor to scalar ratio,   {\it   $r \sim \mathcal{O}(10^{-3}-10^{-2})  $}, with 
observable second order gravitational wave background, which can be {\it potentially} detectable from future experiments, such as DECi-hertz Interferometer Gravitational wave Observatory (DECIGO), Laser Interferometer Space Antenna (eLISA), Cosmic Explorer (CE),  and Big Bang Observatory (BBO).
\end{abstract} 

\maketitle


Detecting the primordial gravitational waves (GWs) will lead to the finest imprints of the nascent Universe, which will confirm the inflationary paradigm~\cite{Inflation}, quantum nature of gravity~\cite{Grishchuk:1974ny,Ashoorioon:2012kh},  and a new scale of physics beyond the Standard Model (BSM). During the slow roll inflation one can excite both scalar and tensor perturbations, see~\cite{Bardeen:1980kt}, and the interesting observable parameter is the tensor-to-scalar ratio, $r$. There are many  models of inflation, see~\cite{Mazumdar:2010sa}, which can predict both large and small $r$, while matching the other observables, such as the amplitude of temperature anisotropy, the tilt in the power spectrum, and its running of the spectrum by the cosmic microwave background radiation (CMBR)~\cite{Ade:2015xua}, within the observed window of ${\cal O}(8)$ e-foldings of primordial inflation from the Planck satellite. However, it is worthwhile also to constrain the potential beyond the 
the {\it pivot scale}, $k_{\ast}= 0.05$~Mpc$^{-1}$, where the relevant observables are normalised.

The aim of this paper will be to provide a simple toy model example of inflationary potential,  which can generate large tensor perturbations, in particular large {\it potentially observable}, $r$, by the ground based experiments such as Bicep-Keck array~\cite{Ade:2015tva}, and also leave imprints of GWs with a frequency  range, $10^{-4}-10^{3}$ Hz, at  DECi-hertz Interferometer Gravitational wave Observatory (DECIGO)~\cite{Decigo}, Laser Interferometer Space Antenna (eLISA)~\cite{AmaroSeoane:2012km}, Cosmic Explorer (CE)~\cite{Evans:2016mbw},  and Big Bang Observer (BBO)~\cite{BBO}, see also~\cite{Moore:2014lga}. 
Therefore, correlating GWs at two different frequencies and wavelengths {\it inspired} by the same model of inflation.

As we will show,  {\it inflection-point} models of inflation~\cite{Allahverdi:2006iq,Mazumdar:2011ih}, provides this unique possibility to excite the GWs from the {\it pivot scale}, where the CMBR observables are normalized  to the end of inflation.

In order to illustrate this,  let us now consider a simple potential which allows {\it inflection-point}, and we will {\it strictly} assume that $\phi_{\rm CMB}, ~\Delta \phi_{\rm CMB}\leq M_p$~\cite{Mazumdar:2011ih,Hotchkiss:2011gz,Chatterjee:2014hna}.
 \begin{equation}
V(\phi) = V_0 + A \phi^2 - B \phi^n + C \phi^{2(n-1)}, 
\label{eq:pot}
\end{equation} 
where $V_0$ corresponds to cosmological constant term during inflation, the coefficients $A,~B,~C$ are appropriate constants with dimensions, and $n\geq 3$ is an integer. The physical motivation for the above potential {\it directly} comes from a softly broken supersymmetric theory with a renormalizable and non-renormalizable superpotential contribution with canonical k\"ahler potential, see~\cite{Allahverdi:2006iq}. In these papers it was assumed that $V_0=0$. However, the supergravity extension, naturally provides  cosmological constant, $V_0$ if no fine tuning is invoked to cancel such a contribution, see for details~\cite{Mazumdar:2011ih}.  Inflation will have to come to an end via phase transition, or via hybrid mechanism~\cite{Linde:1991km}.  In the present work we will also explore the possibility of having large $V_0$, in particular to achieve potentially observable  $r\geq {\cal O}(10^{-3})$ at the pivot scale.

In the above Eq.~(\ref{eq:pot}), $V_0,~A,~B,~C$  are all subject to various cosmological constraints from the latest Planck data~\cite{Ade:2015xua}, here we quote the
central values, which we will use for the reconstruction of $V_0,~A,~B,~C$ from the following {\it well-known} observables:
\begin{align}\label{const-0}
A_\mathrm{s} & \approx  \frac{V}{24 \pi^2 M_\mathrm{pl}^4 \varepsilon_V} \approx2.2\times 10^{-9} \\
n_\mathrm{s}  &\approx 1+2 \eta _V- 6 \varepsilon _V\approx 0.96\label{const-1} \\
\mathrm{d}n_\mathrm{s}/\mathrm{d}\ln k & \approx   16 \varepsilon _V\eta _V- 24 \varepsilon^2_V - 2 \xi^2_V \approx -0.013 \label{const-2} \\
\mathrm{d}^2n_\mathrm{s}/\mathrm{d}\ln k^2  & \approx  -192 \varepsilon^3_V + 192 \varepsilon^2_V \eta _V- 
32 \varepsilon _V\eta^2_V \nonumber \\  
& \quad - 24 \varepsilon _V\xi^2_V + 2 \eta _V\xi^2_V + 2 \sigma^3_V \approx 0.03\label{const-3} \,,
\end{align}
where  $\varepsilon _V, ~\eta _V,~\xi_V,~\sigma_V$ are slow-roll parameters defined below. All the above quantities are measured at the {\it pivot} scale, $k_\ast= 0.05$~Mpc$^{-1}$, and we have considered the central values in this paper, such as $A_{\rm s}(k_\ast)$ is the amplitude of the temperature anisotropy in the CMB, $n_s(k_\ast)$ is the spectral tilt, $\mathrm{d}n_\mathrm{s}/\mathrm{d}\ln k (k_\ast)$ is the running of the tilt and 
$\mathrm{d}^2n_\mathrm{s}/\mathrm{d}\ln k^2(k_\ast)$ designates the running of the the running of the tilt~\cite{Ade:2015xua}. Further note that the slow roll parameters can be expressed in terms of the potential, and given by, see review~\cite{Mazumdar:2010sa}:
\begin{eqnarray}\label{eq:slowroll0}
\varepsilon_V= \frac{M_P^2}{2}\left(\frac{V'}{V}\right)^2; 
~\eta_V=M_P^2 \left(\frac{V''}{V}\right); \\
\xi^2_V=M_P^4 \left(\frac{V'V'''}{V^2}\right);
~\sigma^3_V=M_P^6 \left(\frac{V'^2 V''''}{V^3}\right). 
\label{eq:slowroll}
\end{eqnarray}
Another {\it key} formula is the tensor perturbations and the value of $r$, and its tilt, which are given by:
\begin{eqnarray}
A_\mathrm{t} & \approx & \frac{2 V}{3 \pi^2 M_\mathrm{pl}^4},~~~~r(k=k_\ast) = \dfrac{\mathrm{A}_{t}(k_\ast)}{\mathrm{A}_{s}(k_\ast)}. \label{eq:at_def} \\
n_\mathrm{t} &\approx &- 2 \varepsilon_V, \nonumber
\end{eqnarray}
In fact, the coefficients, $A, ~B, ~C$ can be computed in terms of $V_0$, ${\rm A_s}, r, n_s$, with the help of the following relation, see~\cite{Hotchkiss:2011gz,Chatterjee:2014hna}:
\bea
V(\phicmb) &= &\frac{3}{2} A_s r \pi^2,~~~~
V'(\phicmb) = \frac{3}{2} \sqrt{\frac{r}{8}}(A_s r \pi^2), \nonumber \\
V''(\phicmb) &=& \frac{3}{4} \left(\frac{3r}{8}+n_s-1\right)(A_s r \pi^2). 
\label{eq:recon}
\eea
Given the observable constraints, see Eq.~(\ref{const-0},\ref{const-1},\ref{const-2},\ref{const-3}), we scan the parameter space by fixing the value of $n=3,~4$. 
By insisting that the total number of e-foldings of inflation to be ${\cal N}=50$ along with $\phicmb \sim {\cal O}(M_p)$, we obtain the following benchmark points, as tabulated in Table.~\ref{tab:bm}.

\begin{widetext}
\begin{table}[ht!]
 \begin{center}
  \begin{tabular}{|c|c|c|c|c|c|c|c|c| }
   \hline
Benchmark  & $~~n~~$ & $V_0 (k_\ast) $  &  $A(k_\ast)$  & $B(k_\ast)$  & $C(k_\ast)$ &  $\dfrac{d n_s}{d \ln k}(k_\ast)$ & $~\dfrac{d^2 n_s}{d \ln k^2}(k_\ast)~$   & $~r(k_\ast)~$   \\Points (BP) &  &   &   &   &   &   &    &   \\

 \hline  
 1 & 3  & 7.44$\times 10^{-10}$ & 0.868$\times 10^{-10}$ & 0.689$\times 10^{-10}$ &  0.190 $\times 10^{-10}$ & -0.006 & 0.003  & 0.024    \\
  
 2 & 3  & 1.506$\times 10^{-10} $  &  0.2046 $\times 10^{-10}$& 0.2246$\times 10^{-10}$ & 0.0757$\times 10^{-10}$ & -0.0148 & 0.001   & 0.005 \\ 
 
 3 & 4 & 14.245$\times 10^{-10}$ & 1.240 $\times 10^{-10}$ & 0.500 $\times 10^{-10}$& 0.112 $\times 10^{-10}$& -0.0148 & 0.021 & 0.046     \\   
   \hline
 \end{tabular} 
\caption{We have used $n_s = 0.96,~ A_s = 2.2 \times 10^{-9}, \phi_{\rm CMB} = 1$ in the Planck units for all the benchmarks evaluated at $k_{\ast}= 0.05$~Mpc$^{-1}$. The three benchmark points match the current CMBR data, i.e. the central values used in Eqs.~(\ref{const-0},\ref{const-1}).  } 

\label{tab:bm}
 \end{center}
\end{table}
\end{widetext}

 \begin{figure}[ht!]
\centering
\epsfig{file=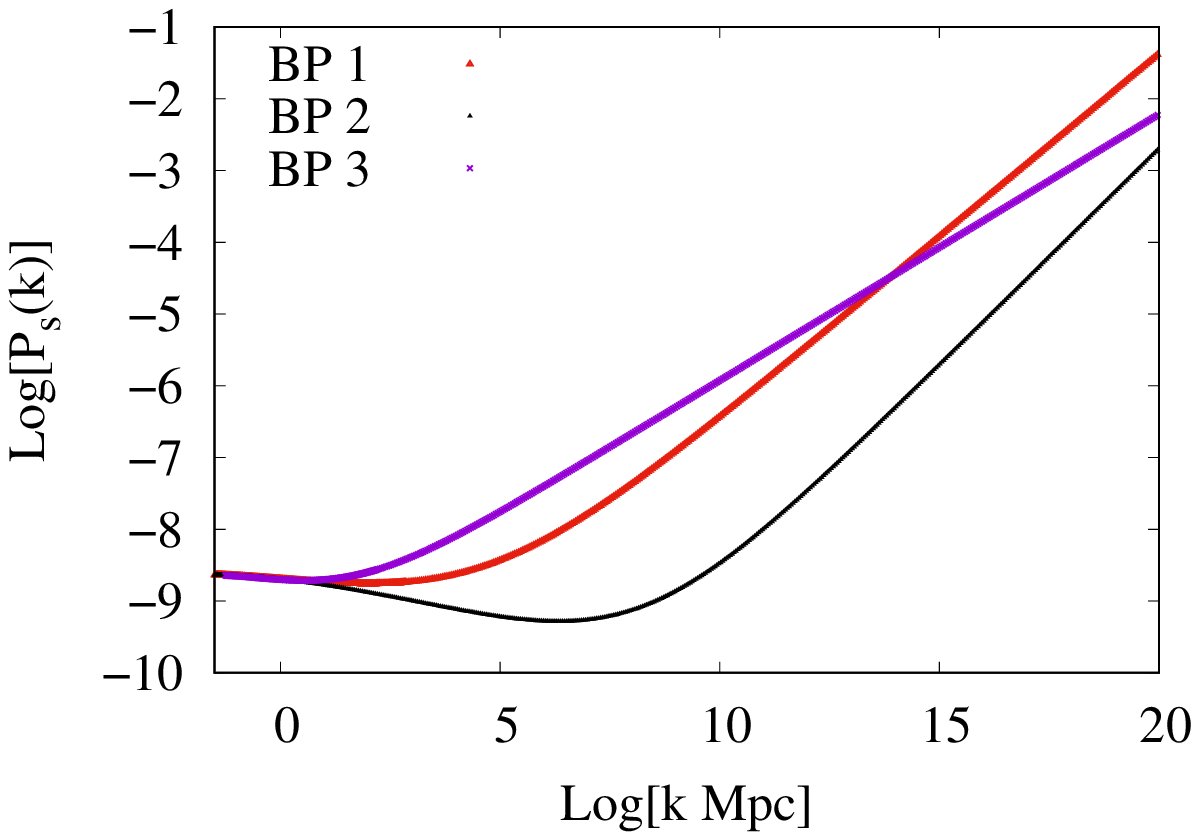,width= \linewidth}
\caption{The scalar power spectra has been shown for the benchmark scenarios in table \ref{tab:bm}.}
  \label{fig:psk}
\end{figure}


We now plot the  amplitude of the scalar power spectrum, $A_s$ in Fig.~[\ref{fig:psk}], for the three benchmark points, see~[\ref{tab:bm}], two of them are for renormalizable potential and one for non-renormalizable potential. We illustrate the power spectrum beyond the Planck window of ${\cal O}(8)$ e-foldings, and show that the scalar amplitude grows outside this 
observable window, and reaches $P_s(k) \leq 10^{-1.5}$ for $k\leq 20{\rm Mpc}^{-1}$ at the end of $50$ e-foldings of inflation. This happens due to the fact that both $\epsilon_V, \eta_V$ change non-monotonically within the observational window of ${\cal O}(8)$ e-foldings. At the {\it pivot } point, $k=0.05 {\rm Mpc}^{-1}$, the scalar power spectrum, the tilt and its running all match the observed data, see Table~\ref{tab:bm}, and Eqs.~(\ref{const-0},\ref{const-1},\ref{const-2},\ref{const-3}), but 
 as soon as the inflaton has crossed  $\phi_{\rm CMB}$, or the {\it pivot} point, the value of $\epsilon_V$ reaches its maximum, and then 
decreases rapidly, while the other slow roll parameter $\eta_V$  decrease before increasing again as $\phi$ decreases ~\cite{Hotchkiss:2011gz,Chatterjee:2014hna}.  At small $\phi \ll \phi_{CMB}$, the slow roll parameter $\eta_V \rightarrow \frac{2 A}{V_0}$. It is the large $\eta_V$ at small $\phi \ll \phi_{CMB}$, that leads to more power at small length scales. This property was first noticed in~\cite{Hotchkiss:2011gz}. 
Note that, for large $V_0$, it can dominate the energy density  well after the CMB observable window to the end of the inflation, inflation will typically end via phase transition as discussed above. 
In our case, there will be a bump-like feature in the potential close to the pivot scale. This, in turn, will give rise to large $r $ corresponding to the benchmark points. In this paper we will
not discuss how to end inflation, and how to reheat the Universe in any detail~\cite{Allahverdi:2010xz}, but we will now ask the possibility of generating GWs at different 
length scales and frequencies.

Now, since the scalar power spectrum has an increasing trend in the infrared, see Fig.~[\ref{fig:psk}], one can ask whether this would
 source any gravitational waves at the second order. The gravitational perturbations can be sourced by the matter perturbations at the second order, this has been studied in 
Refs.~\cite{Ananda:2006af,Baumann:2007zm}. Based on this we can ask how much the amplification of GWs will be at scales around 
${\cal O}(10-20)~{\rm Mpc}^{-1}$? Also, what will be the frequency range of these GWs, and would they be detectable by DECIGO, eLISA, 
CE, and BBO?

In order to understand this amplification of the GWs, let us first study the metric perturbations, defined as, 
\begin{equation}
ds^2 = -a(\eta)^2 [(1+2\Phi)d \eta^2+\{(1-2\Phi)\delta_{ij}+\dfrac{1}{2}h_{ij}\} dx^i d x^j ]\, \nonumber
\end{equation}
where $\Phi$ is the metric potential, we have taken anisotropic stress  to be absent, and $h_{ij}$ denotes the second-order tensor perturbation, which satisfies $h_{i}^i=0,~ h_{i,j}^j=0 $ (i.e. traceless and transverse conditions). We are keen on the tensor perturbations, which can be expressed as follows,  
\begin{equation}
h_{ij}({\bf x}, \eta) = \dfrac{1}{(2 \pi)^{3/2}}\int d^3 {\bf k} e^{i {\bf k.x}}[h_{\bf k}(\eta) e_{ij}({\bf k})+  \overline{h}_{\bf k}(\eta) \tilde{e}_{ij}({\bf k})] \nonumber
\end{equation}
The two polarization tensors in the above equations 
are normalized, such that $e^{ij} e_{ij}=1=\tilde{e}^{ij} \tilde{e}_{ij},  ~ e ^{ij} \tilde{e}_{ij}=0$. 

Note that, at {\it large k ($k \gtrsim 10^{8} ~{\rm Mpc}^{-1}$) of our interest}, the first-order tensor perturbation during inflation is negligible.  By expanding the Einstein tensor and the energy-momentum tensor up to the second-order, and substituting the same in the Einstein equation, the following equation can be obtained \cite{Baumann:2007zm, Ananda:2006af}~\footnote{To compute the power spectrum, and then the corresponding energy density, it is convenient to work in Fourier space. For the `+' polarization $e_{ij}({\bf k})$, The above equation for the tensor perturbations, then, can be recast as, 
The amplitude $\tilde{h}_{\bf k}$, corresponding to the ``$\times$" polarization \footnote{Note that we follow the normalization in \cite{Baumann:2007zm, Ananda:2006af} for the polarization tensors. Several references follow a different convention, see, e.g. ref \cite{Maggiore:1999vm}.} also obeys a similar equation. }, 
\begin{equation}
h_{\bf k}''+2 \mathcal{H} h_{\bf k}'+ k^2 h_{\bf k} = \mathcal{S}({\bf k}, \eta).
\end{equation}
The source term $\mathcal{S}({\bf k}, \eta)$ can be written as \cite{Baumann:2007zm, Ananda:2006af}, 
\begin{eqnarray}
\mathcal{S}({\bf k}, \eta) &= &-4 e^{lm}({\bf k}) \mathcal{S}_{lm}({\bf k})\nonumber\\ &=&\int \frac{d^3 {\bf q} }{(2 \pi)^{3/2}}e^{lm}({\bf k})q_l q_m \mathcal{F}({\bf k},{\bf q},\eta),
\end{eqnarray}
where, 
\begin{eqnarray}
&&\mathcal{F}({\bf k},{\bf q},\eta) = 12 \Phi(q,\eta)\Phi(|{\bf k-q}|,\eta) \\
&&+ \frac{8}{\mathcal{H}} \Phi'(q,\eta)\Phi(|{\bf k-q}|,\eta)+ \frac{4}{\mathcal{H}^2} \Phi'(q,\eta)\Phi'(|{\bf k-q}|,\eta).\nonumber 
\end{eqnarray}


 \begin{figure}[!t]
\centering
  \epsfig{file=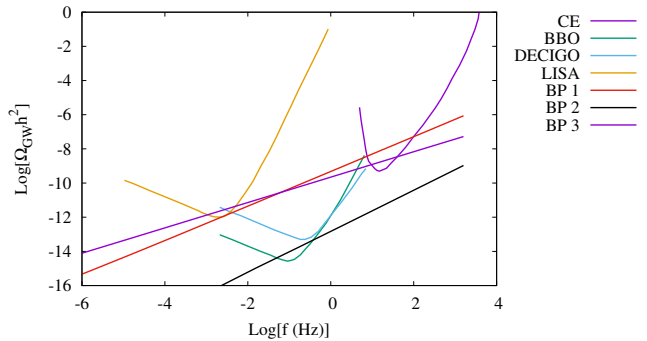,width= \linewidth}
\caption{The relative contribution of the gravitational wave to the energy density  has been shown for the benchmark scenarios in table \ref{tab:bm}.  }
  \label{fig:gw}
\end{figure}%


To estimate the source term, we evaluate the Bardeen potential first~\cite{Bardeen:1980kt}. Since the scalar power spectrum starts rising for $k \gg k_{eq} \sim 0.01 ~{\rm Mpc}^{-1}$, the second-order source term can only be significant for $k \gg k_{eq}$. Consequently, we only consider the modes which are re-entering the Hubble patch during the radiation domination. In this epoch, the Bardeen potential satisfies the following evolution equation : 
\begin{equation} 
\Phi''+ \dfrac{6(1+w)}{(1+3 w) \eta} \Phi'+ w k^2\Phi = 0\,,
\end{equation}
with $w = {1}/{3}$. Ignoring the decaying mode at early times, the   solution takes the following form : 
\begin{equation}
\Phi (k, \eta) = \frac{c(k)}{(k \eta)^3}\left[\frac{k \eta}{\sqrt{3}}\cos\left(\frac{k \eta}{\sqrt{3}}\right) - \sin \left(\frac{k \eta}{\sqrt{3}}\right)\right].
\end{equation}
Note that the Bardeen potential $\Phi (k) $ can be split in to two parts, a contribution from the primordial perturbation $\phi_{\bf k}$ ($\eta \ll 1$) and the transfer function as
$\Phi (k, \eta) = \Phi(k\eta) \phi_{\bf k} $. The coefficient $c(k)$ is estimated 
matching of $\Phi (k, \eta)$ with the primordial perturbation at $\eta \ll 1$. This gives $ \Phi (k, \eta \ll 1)=- {c(k)}/{9 \sqrt{3}}$. Thus $c(k)$ can be estimated from the primordial power spectrum as follows \cite{Ananda:2006af}, 
\begin{equation}
c(k)^2 \simeq (9\sqrt{3})^2\frac{4}{9} \frac{2 \pi^2}{k^3} A_s(k) \simeq \frac{216 \pi^2}{k^3} A_s(k)
\end{equation}
where $A_s(k)$ denote the primordial scalar power spectrum (i.e. the power spectrum as $\eta \rightarrow 0$). Before getting into the numerical results, we describe the behavior of the amplitude $h_{\bf k}$ and the source term first \cite{Baumann:2007zm}.  The amplitude $h_{\bf k}$ is largest at a time $\eta_i$, when $k \eta_i \simeq 1$, i.e. during the period of Hubble re-entry of the respective mode. At this point its amplitude can be simply estimated as ${S({\bf k}, \eta_i)}/{k^2}$. Once a mode enters horizon, it starts  oscillating, and the amplitude decreases as inverse of the scale factor. Also, the source term $S({\bf k})$ decreases  faster during radiation domination before eventually becoming constant during matter dominated epoch. For our benchmarks, see Table~\ref{tab:bm}, we find that the source term scales as 
${1}/{a^{\gamma}}$, where $\gamma \simeq 2-3$. For the modes, which enter  early in the radiation dominated epoch, the source term can become too small before entering the matter dominated epoch, so the amplitude simply decreases as inverse of the scale factor until today. The energy density of the gravitational wave (in logarithmic intervals of $k$) is given by (see e.g. \cite{Maggiore:1999vm}), 
\begin{equation}
\rho_{\rm GW} (k, \eta) = \frac{\langle \dot{h}_{ij} \dot{h}^{ij}\rangle }{32 \pi G}
= \frac{1}{32 \pi G} \frac{k^2}{a(\eta)^2} \mathcal{P}_h(k,\eta),
\end{equation}
where $\eta$ is the conformal time, and the power spectrum $\mathcal{P}_h(k,\eta)$ takes the following form 
\begin{equation}
\mathcal{P}_h(k,\eta) = \frac{k^3}{2 \pi^2} ( |h_{\bf k}(\eta)|^2 +  
|\bar{h}_{\bf k}(\eta)|^2).
\end{equation}
The relative energy density $\Omega_{\rm GW} (k, \eta) = ({1}/{12}) ({k^2}/{a(\eta)^2 H(\eta)^2}) \mathcal{P}_h(k, \eta) $, then, can be estimated at the present epoch  by, $({\Omega_{\rm rad}^0 h^2}/{ \Omega_{\rm rad}^{eq}}) \Omega_{\rm GW}^{eq}(k) $, where we take $h=0.68$, and $\Omega_{\rm GW}^{eq}(k)$ evaluated at the 
re-entry 
\begin{eqnarray}
\Omega_{\rm GW}^0 (k)h^2 
= \frac{ \Omega_{\rm rad}^0 h^2}{  2 \Omega_{\rm rad}^{eq}} \left(\frac{g_{*eq}}{g_{*i}}\right)^{1/3}  \frac{k^2 \mathcal{P}_h(k, \eta_i) }{12 a(\eta_i)^2 H(\eta_i)^2}. 
\end{eqnarray}
where $\eta_i$ represents the conformal time around the Hubble re-entry of the respective mode when the amplitude $h_{\bf k}$ is maximum, thus $k \eta_i \sim \mathcal{O}(1)$. During radiation domination $ \rho_{\rm total} = \rho_{\rm rad} \propto H(\eta)^2 \propto g_{*}^{-1/3} a^{-4}$. Further, the effective number of degree of freedom contributing to the energy density and to the entropy density have been assumed to be the same during this epoch, 
with $g_{*eq} = 106.75,~ g_* = 3.36$ and $\Omega_{\rm rad}h^2 \simeq 4.3 \times 10^{-5}$. We show the estimated $\Omega_{\rm GW}^0 (k)h^2$ for the benchmark scenarios in Fig.\ref{fig:gw}. Note that the BBN and CMBR constraints on $\Omega_{\rm GW}$ (i.e. $\Omega_{\rm GW} \lesssim 10^{-5}$, see e.g. \cite{Smith:2006nka}) is satisfied by our benchmark scenarios. Further, we have also checked that for these scenarios the mass range of primordial blackholes (if they are at all formed due to various astrophysical uncertainties) are typically below $10^{10}$ gm, and therefore no significant constraint arises from their evaporation during early Universe  \cite{Carr:2009jm}. 

Before concluding, let us point out to the key physics for generating large primordial $r$. This is due to the presence of $V_0$ term. It is conceivable that instead of $V_0$, one might be able to invoke many scalar fields giving rise to an enhancement in the Hubble expansion rate ~\cite{Liddle:1998jc}. It would be interesting to see
if multi-scalar fields can also reproduce sufficiently blue tilt in the power spectrum beyond the $8$ e-foldings of observed window via {\it inflection-point} inflation.

To summarise, we have provided an example of inflationary potential, which is capable of generating large tensor-to-scalar ratio, in our scans we 
have given examples of $r=0.024, 0.046, 0.005$. These values of  $r$ are generated by the {\it inflection-point} inflation, which provides large running 
of the slow roll parameters outside the {\it pivot} scale such that the power spectrum increases in the infrared until the end of inflation. The latter sources the secondary 
GWs with $\Omega_{GW}h^2\leq 10^{-6}$, which can be potentially detectable by DECIGO, eLISA, BBO and CE, therefore, opening up new vistas for GW cosmology.

{\it Acknowledgements}: AC acknowledges financial support from the Department of Science and Technology, Government of India through the INSPIRE Faculty Award /2016/DST/INSPIRE/04/2015/000110. 


\end{document}